\documentclass{elsart}
\usepackage{graphicx}
\usepackage{amssymb}

\begin{document}
\begin{frontmatter}
 
\title{Monte Carlo Hamiltonian from Stochastic Basis}

\author[ZSU]{C.Q. Huang},
\author[LU]{H. Kr\"oger\corauthref{cor}},
\author[CCAST,ZSU]{X.Q. Luo},
\author[DU]{K.J.M. Moriarty}

\address[ZSU]{Department of Physics,
       Zhongshan University, Guangzhou 510275, 
       China}

\address[LU]{D\'epartement de Physique, Universit\'e Laval, 
Qu\'ebec, Qu\'ebec G1K 7P4, Canada}

\address[CCAST]{CCAST (World Laboratory), P.O. Box 8730, 
Beijing 100080, China}

\address[DU]{Department of Mathematics, 
Statistics and Computational Science, 
Dalhousie University, Halifax, Nova Scotia B3H 3J5, Canada}

\corauth[cor]{Corresponding author. E-mail: hkroger@phy.ulaval.ca}

\begin{abstract}
In order to extend the recently proposed Monte Carlo Hamiltonian 
to many-body systems,  we suggest to concept of a stochastic basis. 
We apply it to the chain of $N_s=9$ coupled anharmonic oscillators. 
We compute the spectrum of excited states in a finite energy window 
and thermodynamical observables free energy, average energy, 
entropy and specific heat in a finite temperature window. 
Comparing the results of the Monte Carlo Hamiltonian with 
standard Lagrangian lattice calculations, 
we find good agreement. However, the Monte Carlo Hamiltonian results 
show less fluctuations under variation of temperature. 
\end{abstract}


\end{frontmatter}

\section{Introduction}
\label{sec:Introduction}
Path integral quantization
in the Lagrangian formulation and canonical quantization
in the Hamiltonian formulation are two ways
to quantize classical systems. 
The Lagrangian formulation is suitable for numerical simulations 
on a computer via Monte Carlo. 
The enormous success of lattice gauge theory 
over the last two and half decades is  
due to the fact that the Monte Carlo method 
with importance sampling is an excellent technique 
to compute high dimensional (and even ``infinite" dimensional) 
integrals.

Unfortunately, using the Lagrangian formulation 
it is difficult to estimate wave functions and 
the spectrum of excited states. 
Wave functions in conjunction with the energy spectrum 
contain more physical information than the energy spectrum alone. 
Although lattice $QCD$ simulations in the Lagrangian formulation
give good estimates of the hadron masses,
one is yet far from a comprehensive understanding of
hadrons. Let us take as example a new type of hardrons made of
gluons, the so-called glueballs. 
Lattice QCD calculations\cite{kn:Luo96}
predict the mass of the lightest glueball 
with quantum number $J^{PC}=0^{++}$,
to be $1650 \pm 100 MeV$. Experimentally, there are at least two
candidates: $f_0(1500)$ and $f_J(1710)$.
The investigation of the glueball production and decays  
can certainly provide additional important information for
experimental determination of a glueball. 
Therefore, it is important to be able to compute the glueball wave function.

In the Hamiltonian formulation,
one can obtain the ground state energy, 
but also wave functions and the spectrum of excited states.
Often, and in particular in the case of many-body systems, 
it is difficult to solve the stationary Schr\"odinger equation. 
In Ref.\cite{kn:Jirari99} we have suggested how to construct an effective Hamiltonian via Monte Carlo, 
which allows to compute the low energy spectrum 
and the corresponding wave functions. The method has been tested 
in quantum mechanics in D=1,2 and 3 dimensions, for the free system, 
the harmonic oscillator and a number of other local potentials. 
In all cases, the exact results were well reproduced \cite{kn:MCHamilton}.

If one wants to construct the Monte Carlo Hamiltonian in the case of high dimensional systems 
or for many-body systems using a regular basis like in quantum mechanics, one runs into serious troubles. 
Consider, e.g. a 1-D chain of $N$ spin $1/2$ particles. For $N=25$, the dimension of Hilbert space is $d=33554432$, 
which is prohibitively large for numerical computations. 
A physical solution of this problem is the construction of a lower dimensional space 
but retaining the important degrees of freedom.
This principle is realized, e.g., is the folded diagram technique \cite{kn:Kluo71}. 
In recent years, Monte Carlo methods have been widely used to solve problems in quantum physics. 
For example, with quantum Monte Carlo there has been improvement in nuclear shell model calculations \cite{kn:Otsuka}. 
A proposal to solve the sign problem in Monte Carlo Greens function method, 
useful for spin models has been made by Sorello \cite{kn:Sorello}.
Lee et al. \cite{kn:Lee} have suggested a method to diagonalize Hamiltonians, 
via a random search of basis vectors having a large overlap with low-energy eigenstates. 
In contrast to that, in this work we construct a basis, matrix elements of the transition amplitude 
and hence the Hamiltonian via the path integral starting from the action.

We suggest an extension of the idea of the Monte Carlo Hamiltonian with the purpose to treat many-body problems: 
We try to mimic the success of the Monte Carlo method with importance sampling when solving Euclidean path integrals 
in the Lagrangian formulation. There one constructs a ``small" number (say 100 - 1000) of representative 
(equilibrium) configurations and computes the expectation value of an observable by summing the observable 
over those configurations. 
In close analogy to those equilibrium configurations we suggest here to use stochastically 
chosen representative basis states (called stochastic basis in the following).   
This will allow us to construct an effective Hamiltonian from transition matrix elements 
between those stochastic basis states. 
The goal is to solve the many-body problem by construction of the effective Hamiltonian 
in such a ``model space". One expects the dimension of the effective Hamiltonian to be in the same order 
as the typical number of configurations (100-1000) used in computing path integrals in lattice field theory.

We will present results showing that the effective Hamiltonian in conjunction with a stochastic basis works 
in quantum mechanical many-body systems: As example we consider a chain of coupled harmonic anharmonic oscillators 
(scalar $\phi^{4}$ model), and compute thermodynamical observables. 
We find that the Monte Carlo Hamiltonian gives quite accurate results for the spectrum in a finite energy window, 
and also for thermodynamical observables like free energy, average energy, 
entropy and specific heat in a finite temperature window.

\section{Effective Hamiltonian}
\label{sec:Effective Hamiltonian}
Let us briefly recall the meaning of the Monte Carlo Hamiltonian \cite{kn:Jirari99}. 
Using Feynman's path integral formulation \cite{kn:Feynman65}, 
we consider the transition amplitude in imaginary time 
from $t=0$ to $t=T$.  
Using imaginary time makes the path integral mathematically well defined, 
and renders it amenable to numerical simulations by Monte Carlo.
Because the effective Hamiltonian is time independent, 
its construction in imaginary time should give the same result as in real time.
We consider the transition amplitudes for transitions between position states.
Let $\{x_{1},\dots,x_{N}\}$ denote a discrete set of points. 
Then $\{ |x_{1}\rangle, \dots, |x_{N}\rangle \}$ forms a basis of position states. 
We consider the transition amplitudes 
\begin{eqnarray}
M_{ij}(T)
&=& <x_{i} | e^{-H T/\hbar} | x_{j}>
\nonumber \\
&=& 
\int [dx] \exp[ -S_{E}[x]/\hbar ]\bigg |_{x_{j},0}^{x_{i},T} ~ , 
i,j=1,\dots,N ~ ,
\label{eq:DefHeff}               
\end{eqnarray}
where $S_E$ denotes the Euclidean action for a given path $x(t)$ 
going from 
$x_j$, $t=0$ to $x_i$, $t=T$,
\begin{eqnarray}
S_{E}[x(t)] =  S_{0} + S_{V} = 
\int_{0}^{T} dt ~ \frac{1}{2} m \dot{x}^{2} + V(x) ~ . 
\end{eqnarray}
The numerical computation of the matrix elements $M_{ij}$ can be done using standard Monte Carlo 
with importance sampling, by writing it as a ratio of two path integrals, 
thus expressing it as a generalized expectation value of 
an observable $O = \exp[ -S_{V}/\hbar]$. 
\begin{eqnarray}
\label{Matrix_Elements}
M_{ij}(T) &=& M^{(0)}_{ij}(T) ~
\frac{ 
\left. 
\int [dx] ~ \exp[ - S_{V}[x]/\hbar ] ~ \exp[ -S_{0}[x]/\hbar ] \right|^{x_{i},T}_{x_{j},0} }
{ \left.
\int [dx] ~ \exp[ -S_{0}[x]/\hbar ] \right|^{x_{i},T}_{x_{j},0} } 
\nonumber \\
&=& 
M^{(0)}_{ij}(T) ~ \langle  \exp[ - S_{V}[x]/\hbar ] \rangle ~ .
\end{eqnarray}
The matrix elements $M^{(0)}_{ij}$, corresponding to the free action $S_{0}$, 
are known analytically. 
The transition amplitudes $M_{ij}(T)$
form a matrix
$M(T)=[M_{ij}(T)]_{N \times N}$. 
This matrix $M(T)$ is a positive, Hermitian matrix. 
It can be factorized into a unitary matrix $U$ 
and a real diagonal matrix $D(T)$, such that
\begin{eqnarray}
\label{eq:DecompM}
M(T)=U^{\dagger}D(T)U.
\label{eq:MUD}
\end{eqnarray}
Then from Eq.(\ref{eq:DefHeff}), Eq.(\ref{eq:MUD}) one can identify
\begin{eqnarray}
\label{eq:EigenValVec}
U^{\dagger}_{ik}=<x_i|E_k^{eff}>, ~~ D_k(T)=e^{-{E_k^{eff}}T/\hbar}.
\label{eq:UD}
\end{eqnarray}
The $k-th$ eigenvector $|E_k^{eff}>$ 
can be identified with the $k-th$ column of matrix $U^{\dagger}$.
The energy eigenvalues $E^{eff}_{k}$ are obtained from the logarithm 
of the diagonal matrix elements of $D(T)$.
This yields an effective Hamiltonian, 
\begin{eqnarray}
H_{eff} = \sum_{k =1}^{N} | E^{eff}_{k} > E^{eff}_{k} < E^{eff}_{k} |.
\end{eqnarray}

\section{Regular basis}
Above we have not specified the distribution of the nodes $\{x_{1}, \dots, x_{N}\}$. 
Suppose they are equidistantly distributed over the real axis ($x_{i+1} - x_{i} = \Delta x = const$).
Also  we have been a bit too cavalier in using position states $|x_i\rangle$, 
which are not normalizable. To be rigorous, we need to use normalizable states. 
This can be done by constructing ``box"-states, which are 
normalized and localized states.
Those basis states are denoted by $| e_{i} \rangle$, $i=1,\dots,N$. 
They are defined in position space by 
$\langle x|e_{i}\rangle =1/\sqrt{\Delta x}$ in the interval 
$I_{i} = [x_{i} - \frac{1}{2}\Delta x, x_{i} + \frac{1}{2}\Delta x]$, and zero else. 
Hence we consider the transition amplitudes
\begin{eqnarray}
\label{eq:DefMatrix}
M_{ij}(T) 
&=& 
<e_i,T \vert e_j,0>
\nonumber \\
&=& 
\frac{1}{\Delta x} \int_{I_{i}} dx' \int_{I_{j}} dx'' 
\int { [dx] ~ exp [ -S [x] / \hbar ] \bigg{|}^{x',T}_{x'',0} }
\nonumber \\
&=& 
\Delta x 
\int { [dx] ~ exp [ -S [x] / \hbar ] \bigg{|}^{x_i,T}_{x_j,0} } + O(\Delta x^{2}), ~~~  i,j{\in}1,2,\ldots,N.
\end{eqnarray}
Those ``localized" and normalized states, corresponding to equidistantly distributed nodes, 
are called a regular basis.

\section{Stochastic basis}
It is evident that the above basis construction becomes prohibitively large 
if we intend to apply this to a system with many degrees of freedom (many-body system).
For such situations we desire to construct a small basis which gives an effective Hamiltonian 
and reproduces well observables in a low-energy window. 
Why should such a basis exist?
The heuristic argument is the Euclidean path integral, which, when evaluated 
via Monte Carlo with importance sampling, gives a good answer for the 
transition amplitude. This is possible by taking into account a ``small" number of configurations 
(e.g. in the order of 100 - 1000). Roughly speaking, the configurations correspond to basis functions. 
Thus we expect that suitably chosen basis functions exist, the number of which is in the order of 100 - 1000, 
which yields a satisfactory effective low energy Hamiltonian.
Note, however, that this will be the case only when the basis functions are chosen in the ``right" way.

How can we construct such a ``small" basis? Let us consider first the most simple case, 
i.e., a free particle in $D=1$ dimension. Let us take as ``large" basis the regular basis, described above.
Suppose $N$ is large ($N >> 1$). The idea is to make a selection guided by the  
Euclidean quantum mechanical transition amplitude. Recall: For the free system it reads
\begin{eqnarray}
G_{Eucl}(x,T;y,0) = \sqrt{ \frac{m}{2 \pi \hbar T} } 
\exp[ - \frac{m}{2 \hbar T} (x - y)^{2} ] .
\end{eqnarray}
Note that this function is positive for all $x$, $y$, $T$. It can be used as a probability density.
We put $y=0$ and define a probability density $P(x)$ by
\begin{eqnarray}
\label{eq:Probability}
P(x) &=& \frac{1}{Z} G_{Eucl}(x,T;0,0),
\nonumber \\
Z &=& \int dx ~ G_{Eucl}(x,T;0,0) .
\end{eqnarray}
Then we define a selection process as follows: Using a random process 
with probability density $P(x)$ one draws a ``small" set of samples  
$\{ x_{\nu}| \nu \in 1, \dots, N_{stoch} \}$, where $N_{stoch} << N$.

In the case of the free particle, $P(x)$ is a Gaussian,
\begin{eqnarray}
P(x) = \frac{1}{\sqrt{2 \pi} \sigma} \exp [ - \frac{x^{2}}{2 \sigma^{2}} ], ~~
\sigma = \sqrt{ \frac{\hbar T}{m} } .
\label{eq:Gaussian}
\end{eqnarray}
In other words, we select $\{x_{\nu}\}$ by drawing from a Gaussian distribution. We draw $N_{stoch}$ samples, giving $N_{stoch}$ states, where $N_{stoch}$ is considerably smaller than $N$, the original size of the basis.

Let us give some thought to the question: Is such probability density physically reasonable?
Firstly, consider the case when $T$ is large. The Boltzmann-Gibbs distribution
\begin{eqnarray}
P_{BG}(E) = \frac{1}{Z} \exp[ - E ~ T/\hbar ]
\end{eqnarray}
projects onto the ground state when $T \to \infty$. 
For the free system the ground state energy is $E=0$.
I.e., the distribution $P_{BG}(E)$ has a strong peak at $E=0$ (when $T \to \infty$). 
On the other hand, when $T \to \infty$, then $\sigma$ given by Eq.(\ref{eq:Gaussian}) is large. 
Thus the density $P(x)$, from which we draw
the $x_{\nu}$ is a broad Gaussian. In the limit $\sigma \to \infty$, it becomes a uniform distribution.
Now we go over from $P(x)$ to $\tilde{P}(k)$, related via Fourier transformation.
If $P(x)$ is uniform, then $\tilde{P}(k) \propto \delta(k)$.
Thus it gives the energy $E_{k}=\frac{k^{2}}{2m}|_{k=0} = 0$, which is the correct ground state energy eigenvalue. 
Thus, in the extreme low-energy regime, the distribution $P(x)$ 
gives a result consistent with the Boltzmann-Gibbs distribution. 
This is a good indication that $P(x)$ will generate 
an effective Hamiltonian useful for the computation of thermodynamical observables.

Although less relevant for our purpose, it is instructive to 
look what happens in the opposite situation, i.e., when $T$ is small. 
In the limit $T \to 0$, the Boltzmann-Gibbs distribution is approximately a constant. 
All energies occur with equal probability.
Thus $\sigma$ is also small. The distribution $P(x)$ behaves like
$P(x) \propto \delta(x)$. The Fourier transform yields $\tilde{P}(k) =$ const., i.e. a uniform distribution. 
Then $E_{k}=\frac{k^{2}}{2m}$ is distributed like 
$1/\sqrt{E}$. This is not the same as the Boltzmann-Gibbs distribution. But for small $T$, 
which means large energy $E$, it is qualitatively comparable to that of Boltzmann-Gibbs.

Next we ask: What do we do in the case when a local potential is present? 
The definition of the probability density $P(x)$ given by Eq.(\ref{eq:Probability})
generalizes to include also local potentials.
In order to construct a stochastic basis one can proceed via the following steps:
(i) Compute the Euclidean Green's function $G_{E}(x,t;0,0)$, e.g., 
by solving the diffusion equation and compute $P(x)$. 
(ii) Find an algorithm giving a random variable $x$ distributed according to the probability density $P(x)$ 
and draw samples from this distribution, giving nodes, say $x_{\nu}$. 
Finally, one obtains the stochastic basis by constructing 
the corresponding characteristic states from the nodes $x_{\nu}$.

The same goal can be achieved in an elegant and efficient manner via 
the Euclidean path integral. Writing Eq.(\ref{eq:Probability}) as path integral yields
\begin{eqnarray}
\label{eq:ProbDensPathInt}
P(x) = \frac{ \int [dy] \exp[ - S_{E}[y]/\hbar ]\bigg|_{0,0}^{x,T} }
{ \int_{-\infty}^{+\infty} dx 
\int [dy] \exp[ - S_{E}[y]/\hbar ]\bigg|_{0,0}^{x,T} } ~~~ .
\end{eqnarray}
Using a Monte Carlo algorithm with importance sampling, like the Metropolis algorithm \cite{kn:Metropolis53}, 
one generates representative paths, which all start at $x=0$, $t=0$ and arrive 
at some position $x$ at time $t=T$.
Let us denote those paths (configurations) by $C_{j} \equiv x_{j}(t)$.
We denote the endpoint of path $C_{j}$ at time $t=T$ by 
$x^{sto}_{j} \equiv x_{j}(T)$.
Those form the stochastically selected nodes, which define the stochastic basis.

Like we did above for the regular basis, we construct square integrable box states, 
localized around the stochastic nodes.
Those states are denoted by $| e^{sto}_{i} \rangle$, $i=1,\dots,N$. In position space they are defined by
$\langle x|e^{sto}_{i}\rangle =1/\sqrt{\Delta x^{sto}_{i}}$ for $x \in 
I^{sto}_{i} = [x^{sto}_{i} - \frac{1}{2}\Delta x^{sto}_{i}, 
x^{sto}_{i} + \frac{1}{2}\Delta x^{sto}_{i}]$, and zero else. Hence we consider the transition amplitudes
\begin{eqnarray}
\label{eq:DefMatrixStoch}
M_{ij}(T) 
&=& 
<e^{sto}_i,T \vert e^{sto}_j,0>
\nonumber \\
&=& 
\frac{1}{\sqrt{\Delta x^{sto}_{i} \Delta x^{sto}_{j}}} \int_{I^{sto}_{i}} dx' \int_{I^{sto}_{j}} dx'' 
\int { [dx] ~ exp [ -S [x] / \hbar ] \bigg{|}^{x',T}_{x'',0} }
\nonumber \\
&=& 
\sqrt{\Delta x^{sto}_{i} \Delta x^{sto}_{j}} 
\int { [dx] ~ exp [ -S [x] / \hbar ] \bigg{|}^{x^{sto}_i,T}_{x^{sto}_j,0} } 
\nonumber \\
&+& O((\Delta x^{sto})^{2}), 
~~~i,j{\in}1,2,\ldots,N ~ .
\end{eqnarray}

The above expression involves $\Delta x^{sto}_{i}$, i.e. the volume of the interval $I^{sto}_{i}$. 
In a 1-dimensional integral the intervals $I^{sto}_{i}$ have to be chosen such that they cover
the domain filled by the stochastic nodes. However, in higher dimensions this will be complicated. How should one choose then the volume of such ``interval"?
This can be answered by recalling how to compute an integral via Monte Carlo with importance sampling. 
\begin{eqnarray}
J = \int_{a}^{b} dx ~ P(x) ~ g(x) 
\approx  \sum_{i=1}^{N} \Delta x_{i} ~ P(x_{i}) ~ g(x_{i}) ~ .
\end{eqnarray}
Suppose the nodes of integration $x_{i}$ are drawn from the distribution $P(x)$, normalized to unity.  
Then the Monte Carlo estimator of $J$ is given by
\begin{eqnarray}
J_{est} = \frac{1}{N} \sum_{i=1}^{N} g(x_{i}) ~ .
\end{eqnarray}
This tells us that the volume of the ``intervals" is given by
\begin{eqnarray}
\Delta x_{i} = \frac{1}{N} \frac{1}{P(x_{i})} ~ .
\end{eqnarray}
The result holds in arbitrary dimensions.

\section{The model: chain of coupled anharmonic oscillators}
We consider a one-dimensional chain of $N_{s}$ coupled harmonic oscillators, 
with anharmonic perturbation. Its Euclidean action is given by
\begin{eqnarray}
\label{action_osc}
S = \int dt \sum_{n=1}^{N_{s}} ~ \frac{1}{2} \dot{\phi}_{n}^{2}
+ \frac{\Omega^{2}}{2} (\phi_{n+1} - \phi_{n})^{2} 
+ \frac{\Omega_{0}^{2}}{2} \phi_{n}^{2} 
+ \frac{\lambda}{2} \phi_{n}^{4} ~ .
\end{eqnarray}
In the continuum formulation it corresponds to the scalar $\Phi^{4}_{1+1}$ model,  
\begin{eqnarray}
S = \int dt \int dx ~
\frac{1}{2} (\frac{\partial \Phi}{\partial t})^{2} 
+ \frac{1}{2} (\nabla_{x} \Phi)^{2} 
+ \frac{m^{2}}{2} \Phi^{2}
+ \frac{g}{4!} \Phi^{4} ~ .
\end{eqnarray}
Introducing a space-time lattice with lattice spacing $a_{s}$ and $a_{t}$, this action becomes
\begin{eqnarray}
\label{action_scalar}
S &=& \sum_{n=1}^{N_{s}} \sum_{k=0}^{N_{t}-1} a_{t} a_{s} 
\left[ 
\frac{1}{2} \left( \frac{ \Phi(x_{n},t_{k+1}) - \Phi(x_{n},t_{k}) }{a_{t}} \right)^{2}  \right.
\nonumber \\
&+&  \left. \frac{1}{2} \left( \frac{ \Phi(x_{n+1},t_{k}) - \Phi(x_{n},t_{k}) }{a_{s}} \right)^{2}
+ \frac{m^{2}}{2} \Phi^{2}(x_{n},t_{k})
+ \frac{g}{4!} \Phi^{4}(x_{n},t_{k}) \right] ~ .
\end{eqnarray}
The actions given by Eq.(\ref{action_scalar}) and Eq.(\ref{action_osc})
can be identified by posing $\phi=\sqrt{a_s}\Phi$, $\Omega=1/a_s$,
$\Omega_0=m$, and $\lambda/2=g/4!$.

\subsection{Estimation of statistical errors}
\label{sec:Errors}
We have computed the transition matrix elements via Monte Carlo. This yields
matrix elements $M_{ij}$ with statistical errors $\delta M_{ij}$. Using stationary perturbation theory to lowest order, one can compute the propagation of the statistical errors into the energy eigenvalues and wave functions 
of the effective Hamiltonian. Here we have estimated the error propagation numerically. We have considered the matrix $M_{ij} \pm \delta M_{ij}$ and diagonalized it and computed correspondingly $H_{eff}$ via Eqs.(\ref{eq:DecompM},\ref{eq:EigenValVec}). This gives upper bounds on the error of energy eigenvalues and on the error in the wave functions.

\subsection{Spectrum}
In Tab. \ref{tab.1} we present the energy spectrum.
%
%
%
%
%
\begin{table}[tbp]
\caption{Energy spectrum from MC Hamiltonian.
Model parameters: $\Omega=1$, $\Omega_0=2$,   
and $\lambda=1$ ($\hbar=1$, $k_B=1$). 
Approximation parameters: $N_s=9$, $a_s=1$, $\beta=2$.}
\vspace{3mm}
\begin{center}
\begin{tabular}{|c|c|c|c|c|c|}
\hline
$n$ & $E_{n}^{\rm{eff}}$  & statistical error & $n$ & $E_{n}^{\rm{eff}}$  & statistical error \\
\hline
 1  & 11.278101  & 0.013907 & 29 & 15.752267  & 0.068596 \\
 2  & 13.412064  & 0.035563 & 30 & 15.846867  & 0.046377 \\
 3  & 13.566906  & 0.034268 & 31 & 15.875840  & 0.021624 \\
 4  & 13.711700  & 0.044635 & 32 & 15.912447  & 0.079069 \\
 5  & 13.729973  & 0.040317 & 33 & 15.956051  & 0.065853 \\
 6  & 13.877653  & 0.022487 & 34 & 15.965277  & 0.080568 \\
 7  & 14.043190  & 0.019916 & 35 & 16.169730  & 0.119116 \\
 8  & 14.093273  & 0.001820 & 36 & 16.220694  & 0.197029 \\
 9  & 14.255088  & 0.023932 & 37 & 16.222774  & 0.049391 \\
 10 & 14.256074  & 0.024682 & 38 & 16.324841  & 0.071902 \\
 11 & 14.383945  & 0.006369 & 39 & 16.409331  & 0.064214 \\
 12 & 14.450105  & 0.003479 & 40 & 16.520363  & 0.222361 \\
 13 & 14.529230  & 0.041190 & 41 & 16.634297  & 0.074976 \\
 14 & 14.649623  & 0.090273 & 42 & 16.635276  & 0.099756 \\
 15 & 14.723430  & 0.007810 & 43 & 16.850276  & 0.124949 \\
 16 & 14.793013  & 0.035949 & 44 & 16.869804  & 0.578303 \\
 17 & 14.812091  & 0.031060 & 45 & 17.276424  & 0.051881 \\
 18 & 14.904202  & 0.012391 & 46 & 17.373372  & 0.346835 \\
 19 & 15.139851  & 0.068791 & 47 & 17.473837  & 0.613308 \\
 20 & 15.147149  & 0.024694 & 48 & 17.658680  & 0.392987 \\
 21 & 15.159335  & 0.008713 & 49 & 17.783055  & 0.579326 \\
 22 & 15.321950  & 0.031130 & 50 & 18.492900  & 0.445111 \\
 23 & 15.348790  & 0.160528 & 51 & 18.562615  & 0.010174 \\
 24 & 15.367387  & 0.037622 & 52 & 18.716123  & 0.630255 \\
 25 & 15.514476  & 0.040735 & 53 & 18.800322  & 0.063546 \\
 26 & 15.534130  & 0.047999 & 54 & 19.228190  & 0.691547 \\
 27 & 15.585427  & 0.060088 & 55 & 19.585549  & 0.129330 \\
 28 & 15.619200  & 0.017032 & 56 & 19.968143  & 2.334383 \\
\hline
\end{tabular}
\end{center}
\vspace{0mm}
\label{tab.1}
\end{table}

\subsection{Thermodynamical observables}
A solid test of the Monte Carlo Hamiltonian method is a comparison 
with results from standard Lagrangian lattice calculations. 
However, the strength of the latter approach lies not in the computation of excitation spectra.
On the other hand, it does very well for the computation of thermodynamical observables at thermodynamical equilibrium. 
The information of the energy spectrum enters into such thermodynamical functions.
Thus we have chosen to compute the following thermodynamical observables:
the partition function $Z$, free energy $F$,
average energy $U$, specific heat $C$, entropy $S$ and pressure $P$.
Those are defined by
\begin{eqnarray}
Z(\beta) &=& {\rm Tr} \left[ \exp \left( -\beta H \right) \right] ~ ,
\nonumber \\
F(\beta) &=& - \frac{1}{\beta} \log Z ~ ,
\nonumber \\
U(\beta) &=& {1 \over Z} {\rm Tr} \left[ H \exp \left( -\beta H \right) \right]
= - {\partial \log Z \over \partial \beta} ~ ,
\nonumber \\
C(\beta) &=& {\partial U \over \partial \tau}|_{V}
= -k_B \beta^2 {\partial U \over \partial \beta}|_{V} ~ ,
\nonumber \\
S(\beta) &=& \frac{1}{\tau}(U - F) = k_{B} \beta (U - F) ~ ,
\nonumber \\
P(\beta) &=& - {\partial F \over \partial V } ~ .
\end{eqnarray}
Here $k_B$ denotes the Boltzmann constant.  
The temperature $\tau$ is related to $\beta$ via 
$\beta =T/\hbar= 1/({k_B} \tau)$.

\noindent (a) Computation of thermodynamics from the Monte Carlo Hamiltonian: \\
When we approximate $H$ by $H_{\rm{eff}}$,
we can express thermodynamical observables
via the eigenvalues of the effective Hamiltonian
\begin{eqnarray}
\label{eq:MCHThermodyn}
Z^{\rm{eff}}(\beta) &=& \sum_{n=1}^{N} e^{-\beta E_{n}^{\rm{eff}}} ~ ,
\nonumber \\
F^{\rm{eff}}(\beta) &=& - \frac{1}{\beta} \log Z^{\rm{eff}}(\beta) ~ ,
\nonumber \\
U^{\rm{eff}}(\beta) &=& \frac{1}{Z^{\rm{eff}}(\beta)} ~ \sum_{n=1}^{N}
E_{n}^{\rm{eff}} e^{-\beta E_{n}^{\rm{eff}}} ~ ,
\nonumber \\
C^{\rm{eff}}(\beta) &=& k_B \beta^{2} \frac{1}{Z^{\rm{eff}}(\beta)} 
\left[ \sum_{n=1}^{N} (E_{n}^{\rm{eff}})^2 e^{-\beta E_{n}^{\rm{eff}}} 
 - ( \sum_{n=1}^{N} E_{n}^{\rm{eff}} e^{-\beta E_{n}^{\rm{eff}}} )^{2}  
  \right] ~ ,
\nonumber \\
S^{\rm{eff}}(\beta) &=& k_{B} \beta 
\left[ U^{\rm{eff}}(\beta) - F^{\rm{eff}}(\beta) \right] ~ ,
\nonumber \\
P^{\rm{eff}}(\beta) &=& - ~ \frac{ F^{\rm{eff}}(\beta,V + \Delta V) - F^{\rm{eff}}(\beta,V) }{\Delta V} ~ .
\end{eqnarray}
All eigenvalues have been computed from matrix elements, Eq.(\ref{eq:DefMatrix}), in which enters the transition time $T$ or the corresponding value of $\beta$. 
When going to the continuum limit $a_{s} \to 0$, $a_{t} \to 0$, as well as to the thermodynamic limit (infinite volume limit), the energy spectrum should become independent of the temperature (i.e. of $\beta$ and $\tau$) ("perfect scaling").
In practice we have worked on a finite lattice volume ($N_{s}$ nodes in spatial direction) and also used finite lattice spacing $a_{s}$ and $a_{t}$.
Consequently, the energy spectrum obtained will have a finite volume dependence  and also some dependence on the lattice resolution $a_{s}$ and $a_{t}$.
In practice this manifests itself in the existence of an ``energy window" and a ``temperature window" (``scaling window") where the eigenvalues are close to those of the continuum and infinite volume limit, and consequently depend very little on $\beta$ entering the transition amplitudes. This scaling behavior can be improved and the window can be enlarged by approaching the continuum limit, and the infinite volume limit. It also requires to increase $\beta$ and to improve the statistics in the Monte Carlo computation of matrix elements.  
The numerical analysis of those scaling properties merits a detailed numerical study, which we defer to a future study.
In the numerical results presented below we have computed the spectrum at a fixed transition time $T_{0} = \beta_{0} = 2$ ($\hbar = k_{B}=1$).

One should note that once the energy spectrum has been obtained in such a window, thermodynamic functions can be 
computed easily for all values of $\beta$ in the temperature window, from Eqs.(\ref{eq:MCHThermodyn}).
This property is a nice feature of the Monte Carlo Hamiltonian approach.
It is in contrast to the Lagrangian approach, where all thermodynamic functions 
at a particular value of $\beta$ require an independent simulation.

\noindent (b) Computation of thermodynamics in the standard Lagrangian lattice formulation: \\
First, let us consider the average energy $U(\beta)$.
It is given by the path integral
\begin{eqnarray}
\label{eq:ULattice}
U(\beta) 
&=& 
\frac{ - \frac{\partial}{\partial \beta} 
\frac{1}{K} 
\int [\prod_{k=0}^{N_{t}-1} d\phi_{k}] ~ exp [ -S [\phi] ] \bigg{|}^{\phi_{0},t=\beta}_{\phi_{0},t=0} }
{ \frac{1}{K} 
\int [\prod_{k=0}^{N_{t}-1} d\phi_{k}] ~ exp [ -S [\phi] ] \bigg{|}^{\phi_{0},t=\beta}_{\phi_{0},t=0} } ~ .
\end{eqnarray}
Putting $\beta = N_{t} a_{t}$, differentiation with respect to $\beta$ can be expressed as differentiation with respect to $a_{t}$. 
One obtains
\begin{eqnarray}
U(\beta) 
&=& 
\frac{N_s}{2 a_{t}}
+ \frac{1}{N_{t}}
\frac{ \int [\prod_{k=0}^{N_{t}-1} d\phi_{k}] ~ \frac{\partial}{\partial a_{t}} S[\phi] ~ exp [ -S[\phi] ] \bigg{|}^{\phi_{0},t=\beta}_{\phi_{0},t=0} }
{ \int [\prod_{k=0}^{N_{t}-1} d\phi_{k}] ~ exp [ -S[\phi] ] \bigg{|}^{\phi_{0},t=\beta}_{\phi_{0},t=0} } 
\nonumber \\
&=&
\frac{N_s}{2 a_{t}} + \frac{1}{N_{t}} \langle \frac{\partial}{\partial a_{t}} S \rangle ~ .
\end{eqnarray}

 The computation of the free energy $F$ poses some problem in Lagrangian lattice formulation. 
This has to do with the fact that unlike the average energy $U$, 
$F$ can not easily be written as some expectation value of an observable (like 
$O = \frac{\partial}{\partial a_{t}} S[\phi]$ for average energy).
One possibility is to first compute $U(\beta)$ and then integrate over $\beta$.
However, that gives $F(\beta)$ only up to a constant. 
Another possibility is to write the partition function of the scalar model  
as partition function of the Klein-Gordon model times an expectation value 
involving the $\phi^{4}$ interaction in a path integral with the Klein-Gordon action as weight factor,
\begin{eqnarray}
Z^{\phi^{4}}(\beta) = Z^{KG}(\beta) ~ \langle \exp[ - \frac{\lambda}{2} \phi^{4} ] \rangle_{KG} ~ .
\end{eqnarray}
Then one obtains the free energy from
\begin{eqnarray}
F^{\phi^{4}}(\beta) = F^{KG}(\beta) - \frac{1}{\beta} \log \langle \exp[ - \frac{\lambda}{2} \phi^{4} ] \rangle_{KG}(\beta) ~ .
\label{free_energy_MC}
\end{eqnarray}

The entropy, being essentially a difference between average energy and free energy 
is easily obtained from the former two quantities.

Like the average energy $U(\beta)$ also the specific heat $C(\beta)$ 
can be expressed as a combination of expectation values involving derivatives of the action with respect to $a_{t}$.
\begin{eqnarray}
\label{eq:SpecHeatLagrangian}
C(\beta) 
&=& 
- k_{B} \beta^{2} \left[ - \frac{N_s}{2 N_{t} a_{t}^{2}} 
+ \frac{1}{N_{t}^2} \left[ 
\langle 
\frac{\partial^{2} S}{\partial a_{t}^{2}} 
- (\frac{\partial S}{\partial a_{t}})^{2} 
\rangle 
+ 
\langle \frac{\partial S}{\partial a_{t}} \rangle^{2} 
\right] \right] ~ .
\end{eqnarray}

Let us compare the results from the MC Hamiltonian with those from the standard Lagrangian lattice approach.
We haven chosen $N_s=9$ and $a_s=1$ (note that we have made no attempt to go to the continuum limit of the quantum theory,
but our purpose is to compare both methods on given finite lattice). 
As model parameters we have taken $\Omega=1$, $\Omega_0=2$,   
and $\lambda=1$ ($\hbar=1$, $k_B=1$). In the Monte Carlo Hamiltonian 
simulation we used a stochastic basis of $N_{stoch}=100$ states, and
we used $N_{conf}=300$ configurations to measure the matrix elements. 
In the Lagrangian lattice simulation we used $N_{conf}=100-200$ configurations to measure the free energy $F$, and
$N_{conf}=100 000-200 000$ configurations to measure $U$ and $C$.
It took about 192 h CPU on a Pentium 500 computer to obtain all Monte Carlo Hamiltonian results, while the Lagrangian lattice simulations took 2.8 h CPU for each value of $\beta$.

We found in the Monte Carlo Hamiltonian approach that the matrix elements 
are not very sensitive to the choice of $a_t$, provided that $a_{t}$ is sufficiently small.
This means the discretization error in the path integral is quite small. 
The results presented in the Figs. below correspond to
$a_t=1/30$. In the Lagrangian lattice approach 
$U$ is stable for this value of $a_t$.
However, the results for $F$ and $C$ are very sensitive to the choice of $a_t$.
Note, in Eq.(\ref{free_energy_MC}) for the free energy,
the analytical formula $F^{KG}(\beta)$ is has been used, which corresponds to the limit $a_t \to 0$. 
Thus we have chosen $a_{t}$ to be small, $a_t=0.01$, in this case.
On the other hand, when measuring the specific heat, a small value of $a_t$ leads to very large fluctuations. 
$C$ is stable only
for $a_t \ge 0.1$. Therefore, we have chosen $a_t=0.1$ in that case.

\begin{figure}[thb]
 \begin{center}
 \includegraphics[scale=0.6,angle=270]{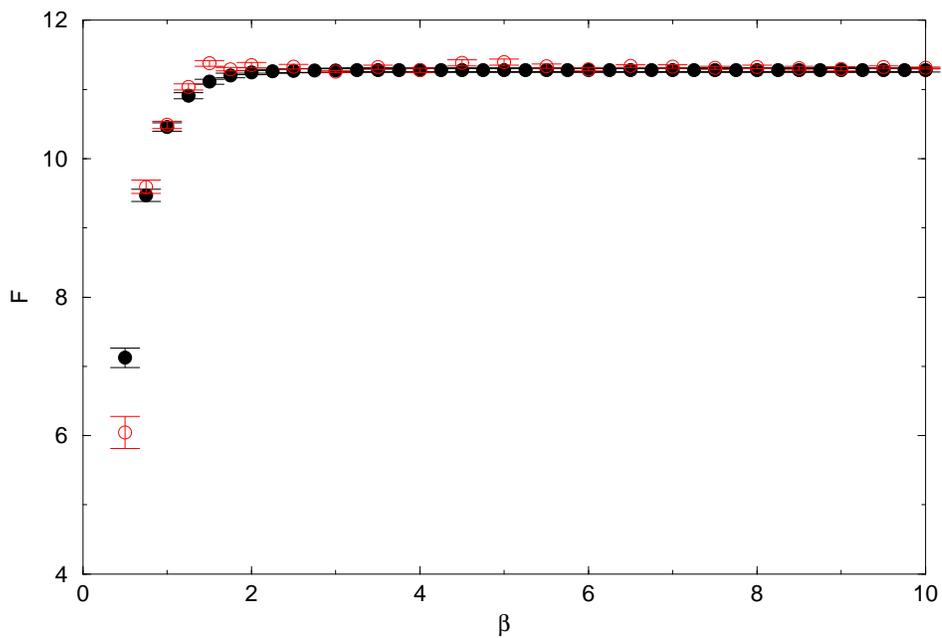}
 \end{center}
 \caption{Free energy $F({\beta})$. 
Comparison of results from Monte Carlo Hamiltonian (filled circles) with standard Lagrangian lattice calculations
(open circles).}
 \label{Fig1}
 \end{figure}

\begin{figure}[thb]
 \begin{center}
 \includegraphics[scale=0.6,angle=270]{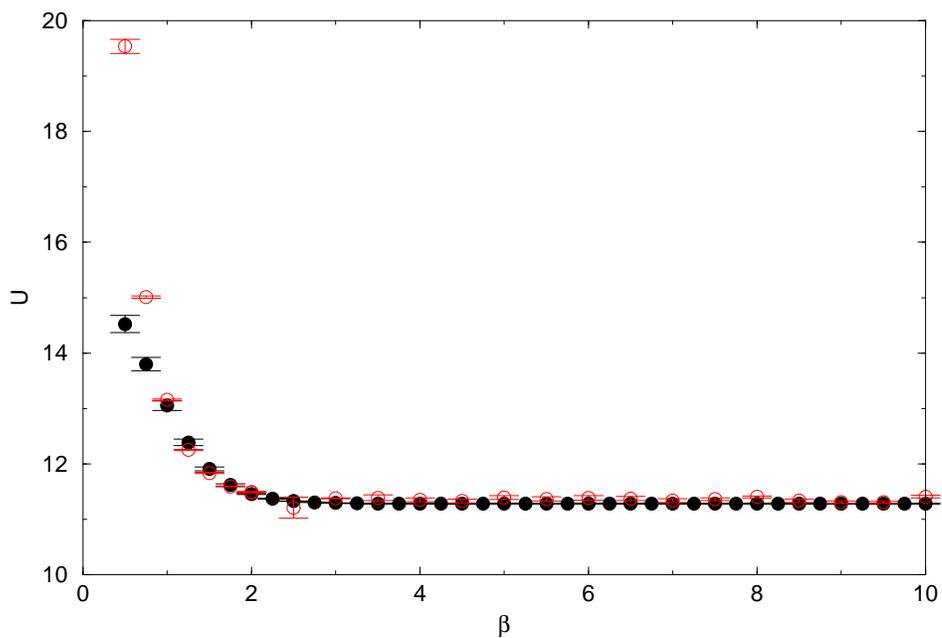}
 \end{center}
 \caption{Same as Fig. \ref{Fig1}, for average energy $U({\beta})$.}
 \label{Fig2}
 \end{figure}

\begin{figure}[thb]
 \begin{center}
 \includegraphics[scale=0.6,angle=270]{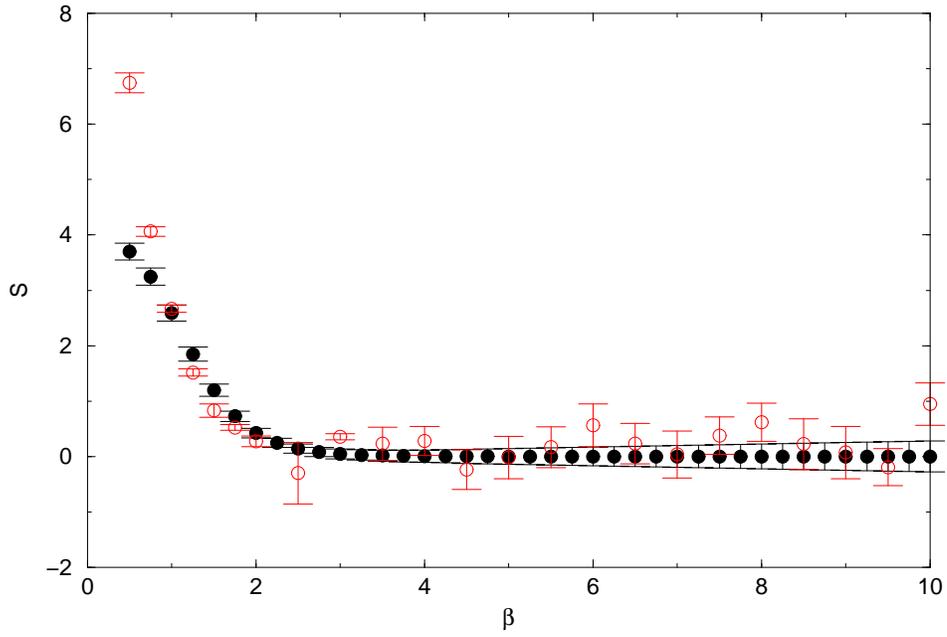}
 \end{center}
 \caption{Same as Fig. \ref{Fig1}, for entropy $S({\beta})$.}
 \label{Fig3}
 \end{figure}

\begin{figure}[thb]
 \begin{center}
 \includegraphics[scale=0.6,angle=270]{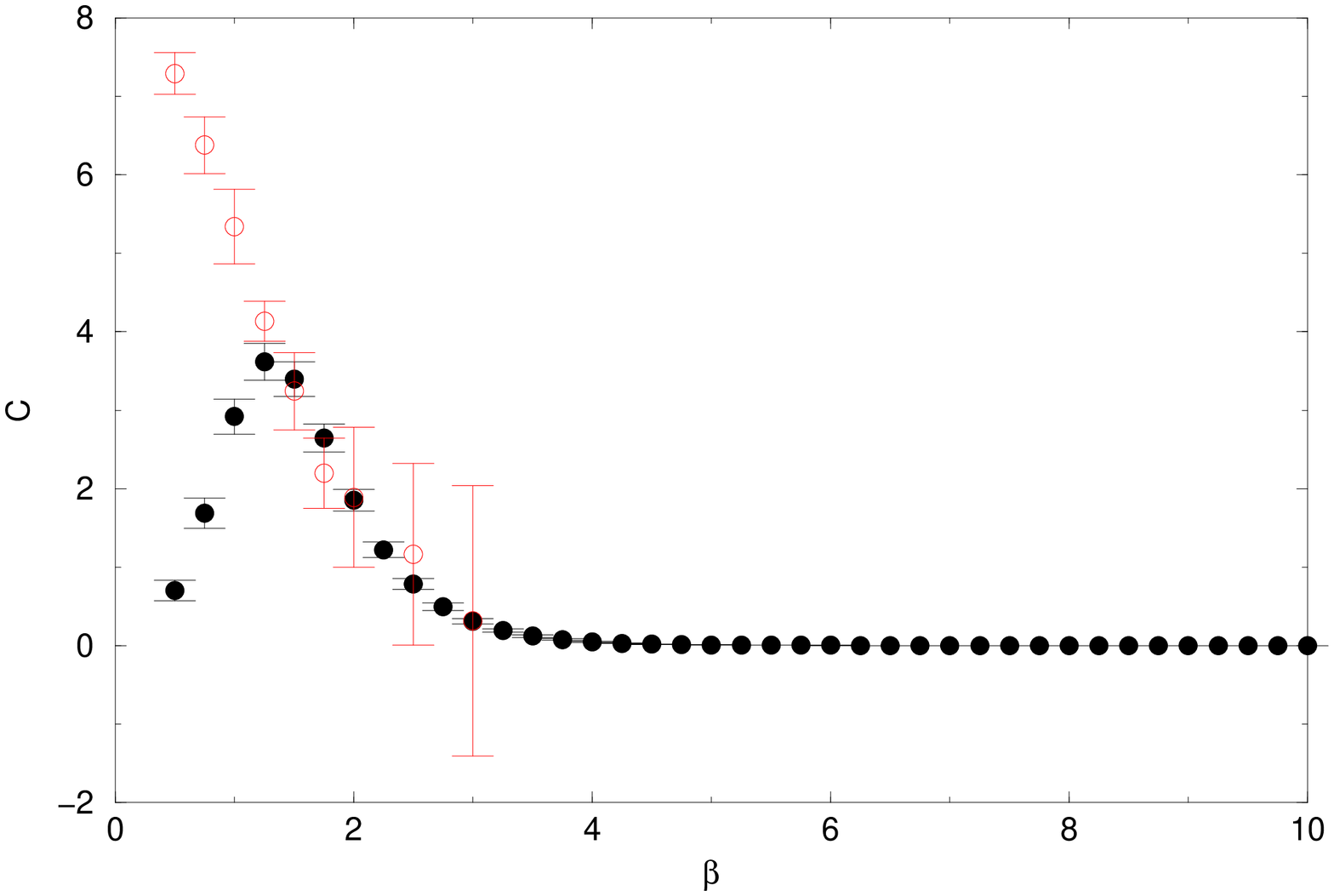}
 \end{center}
 \caption{Same as Fig. \ref{Fig1},  for specific heat $C({\beta})$.}
 \label{Fig4}
 \end{figure}

Fig. \ref{Fig1} shows the free energy $F$ as a function of $\beta$.
We make the following observations: First, there is good overall agreement in the range $1 \le \beta \le 10$.
Second, the Lagrangian lattice data fluctuate more than those from the MC Hamiltonian. 
However, the estimated statistical errors are comparable. 
Fig. \ref{Fig2} shows the average energy $U(\beta)$. 
The behavior is qualitatively the same as for the free energy. 
One notes that at $\beta=0.5$ a marked difference shows up between MC Hamilton and Lagrangian lattice data.
Fig. \ref{Fig3} displays the entropy $S(\beta)$. 
Because the entropy is essentially given by the difference of average and free energy, 
amplified by the factor $\beta$, one observes an amplification in the fluctuations 
of the Lagrangian lattice data when $\beta$ increases (temperature goes to zero). 
For the same reason, also the statistical errors increase. 
However, the MC Hamilton data are stable in that limit. 
Again a marked difference between the two methods become apparent at $\beta=0.5$.
Finally, we display in Fig. \ref{Fig4} the results of the specific heat $C(\beta)$. 
The computation of this function in the Lagrangian lattice approach 
involves second order derivatives and the occurrence of cancellations (see Eq.(\ref{eq:SpecHeatLagrangian})).  
This requires very high statistics, in particular for large $\beta$ ($\beta \ge 3$). 
Hence we have measured $C(\beta)$ only up to $\beta =3$ in the Lagrangian lattice method.  
From the data of $F$, $U$, $S$ and $C$, we estimate the temperature window 
of the Monte Carlo Hamiltonian to range from $\beta=1$ to $\beta=10$.
As discussed above, the size of the window is expected to depend on the model parameters 
as well as on the approximation parameters.

\section{Discussion}
\label{sec:Discussion}
We have suggested how to obtain an effective 
low energy Hamiltonian by constructing via Monte Carlo a 
stochastic basis. We want to stress that the advantage of using a stochastic basis 
shows up in high-dimensional systems and in many-body systems. 
This is based on the experience with integrals, 
where summation over Monte Carlo nodes wins over fixed node rules for dimensions $D>6$, 
as a rule of thumb. We have shown in the scalar model that the MC Hamiltonian 
with stochastic basics works well by computing thermodynamical observables. 
While thermodynamical functions can be obtained also from the standard Lagrangian lattice approach, 
we found that the results from the MC Hamiltonian 
display less fluctuations. Moreover, we have shown that the MC Hamiltonian 
also provides the spectrum and corresponding wave functions in some finite energy window, 
which is very difficult to obtain in the Lagrangian lattice approach. 
It is here, in our opinion, where the MC Hamiltonian approach has an advantage.
Examples of physics where wave functions of many-body systems play a role are:
Hadron structure functions in particle physics, electromagnetic form factors in nuclear physics, 
Bose-Einstein condensation in atomic physics.
We hope that the Monte Carlo Hamiltonian will allow to make progress in those areas.

\ack

H.K. and K.J.M.M. are grateful for support by NSERC Canada.
X.Q.L. is supported by the
National Science Fund for Distinguished Young Scholars,
National Science Foundation of China, 
the Ministry of Education, 
the Foundation of
the Zhongshan University Advanced Research Center
and the Guangdong Provincial Natural Science Foundation of China (proj. 990212).


\begin{thebibliography}{9}

\bibitem {kn:Luo96} X.Q.~Luo and Q.~Chen, Mod. Phys. Lett. A11 (1996) 2435; \\
X.Q.~Luo, Q.~Chen, S.~Guo, X.~Fang, J.~Liu,
Nucl. Phys. B(Proc. Suppl.) 53 (1997) 243.

\bibitem {kn:Jirari99} H.~Jirari, H.~Kr\"oger, X.Q.~Luo and K.J.M.~Moriarty, 
Phys. Lett. A258 (1999) 6.

\bibitem{kn:MCHamilton}
X.Q.~Luo, C. Huang, J. Jing, H.~Jirari, H.~Kr\"{o}ger, K.J.M.~Moriarty, Physica A281 (2000) 201; \\
C.~Huang, J.~Jiang, X.Q.~Luo, H.~Jirari, H.~Kr\"{o}ger, K.J.M.~Moriarty, High Energy Phys. Nucl. Phys. 24 (2000) 478; \\ 
J.~Jiang, C. Huang, X.Q. Luo, H.~Jirari, H.~Kr\"{o}ger, K.J.M.~Moriarty, Commun. Theor. Phys. 34 (2000) 723; \\
X.~Q.~Luo, H.~Xu, J.~Yang, Y.~Wang, D.~Chang, Y.~Lin, H.~Kroger,
Commun. Theor. Phys.  36 (2001) 7.


\bibitem {kn:Kluo71} T.T.S.~Kluo, S.Y.~Lee and K.F.~Ratcliff, 
Nucl. Phys. A176 (1971) 65.

\bibitem{kn:Otsuka} N. Shimizu, T. Otsuka, T. Mizusaki, M. Honma,
Phys. Rev. Lett. 86 (2001) 1171; \\
M. Honma, T. Mitzusaki, T. Otsuka, Phys. Rev. Lett. 77 (1996) 3315.

\bibitem{kn:Sorello} S. Sorello, Phys. Rev. Lett. 80 (1998) 4558.

\bibitem{kn:Lee} D. Lee, N. Salwen, D. Lee, Phys. Lett. B503 (2001) 223.

\bibitem {kn:Feynman65} R.P.~Feynman and A.R.~Hibbs, 
{\it Quantum Mechanics and Path Integrals}, McGraw-Hill, New York (1965).

\bibitem {kn:Metropolis53} N.~Metropolis, A.~Rosenbluth, M.~Rosenbluth, 
A.~Teller, E.~Teller, J. Chem. Phys. 21 (1953) 1087.

\end{thebibliography}
\end{document}